# Reconstruction of Sparse Circuits Using Multi-neuronal Excitation (RESCUME)

**Tao Hu and Dmitri B. Chklovskii**
Janelia Farm Research Campus, HHMI
19700 Helix Drive, Ashburn, VA 20147
*hut, mitya@janelia.hhmi.org*

## Abstract

One of the central problems in neuroscience is reconstructing synaptic connectivity in neural circuits. Synapses onto a neuron can be probed by sequentially stimulating potentially pre-synaptic neurons while monitoring the membrane voltage of the post-synaptic neuron. Reconstructing a large neural circuit using such a "brute force" approach is rather time-consuming and inefficient because the connectivity in neural circuits is sparse. Instead, we propose to measure a post-synaptic neuron's voltage while stimulating sequentially random subsets of *multiple* potentially pre-synaptic neurons. To reconstruct these synaptic connections from the recorded voltage we apply a decoding algorithm recently developed for compressive sensing. Compared to the brute force approach, our method promises significant time savings that grow with the size of the circuit. We use computer simulations to find optimal stimulation parameters and explore the feasibility of our reconstruction method under realistic experimental conditions including noise and non-linear synaptic integration. Multi-neuronal stimulation allows reconstructing synaptic connectivity just from the spiking activity of post-synaptic neurons, even when sub-threshold voltage is unavailable. By using calcium indicators, voltage-sensitive dyes, or multi-electrode arrays one could monitor activity of multiple post-synaptic neurons simultaneously, thus mapping their synaptic inputs in parallel, potentially reconstructing a complete neural circuit.

## 1      Introduction

Understanding information processing in neural circuits requires systematic characterization of synaptic connectivity [1, 2]. The most direct way to measure synapses between a pair of neurons is to stimulate potentially pre-synaptic neuron while recording intra-cellularly from the potentially post-synaptic neuron [3-8]. This method can be scaled to reconstruct multiple synaptic connections onto one neuron by combining intracellular recordings from the post-synaptic neuron with photo-activation of pre-synaptic neurons using glutamate uncaging [9-13] or channelrhodopsin [14, 15], or with multi-electrode arrays [16, 17]. Neurons are sequentially stimulated to fire action potentials by scanning a laser beam (or electrode voltage) over a brain slice, while synaptic weights are measured by recording post-synaptic voltage.

Although sequential excitation of single potentially pre-synaptic neurons could reveal connectivity, such a "brute force" approach is inefficient because the connectivity among neurons is sparse. Even among nearby neurons in the cerebral cortex, the probability of connection is only about ten percent [3-8]. Connection probability decays rapidly with the distance between neurons and falls below one percent on the scale of a cortical column [3, 8]. Thus, most single-neuron stimulation trials would result in zero response making the brute force approach slow, especially for larger circuits.

Another drawback of the brute force approach is that single-neuron stimulation cannot be combined efficiently with methods allowing parallel recording of neural activity, such as calcium imaging [18-22], voltage-sensitive dyes [23-25] or multi-electrode arrays [17, 26]. As these techniques do not reliably measure sub-threshold potential but report only spiking activity, they would reveal only the strongest connections that can drive a neuron to fire [27-30]. Therefore, such combination would reveal only a small fraction of the circuit.

We propose to circumvent the above limitations of the brute force approach by stimulating multiple potentially pre-synaptic neurons simultaneously and reconstructing individual connections by using a recently developed method called compressive sensing (CS) [31-35]. In each trial, we stimulate $F$ neurons randomly chosen out of $N$ potentially pre-synaptic neurons and measure post-synaptic activity. Although each measurement yields only a combined response to stimulated neurons, if synaptic inputs sum linearly in a post-synaptic neuron, one can reconstruct the weights of individual connections by using an optimization algorithm. Moreover, if the synaptic connections are sparse, i.e. only $K << N$ potentially pre-synaptic neurons make synaptic connections onto a post-synaptic neuron, the required number of trials $M \sim K \log(N/K)$, which is much less than $N$ [31-35].

The proposed method can be used even if only spiking activity is available. Because multiple neurons are driven to fire simultaneously, if several of them synapse on the post-synaptic neuron, they can induce one or more spikes in that neuron. As quantized spike counts carry less information than analog sub-threshold voltage recordings, reconstruction requires a larger number of trials. Yet, the method can be used to reconstruct a complete feedforward circuit from spike recordings.

Reconstructing neural circuit with multi-neuronal excitation may be compared with mapping retinal ganglion cell receptive fields. Typically, photoreceptors are stimulated by white-noise checkerboard stimulus and the receptive field is obtained by Reverse Correlation (RC) in case of sub-threshold measurements or Spike-Triggered Average (STA) of the stimulus [36, 37]. Although CS may use the same stimulation protocol, for a limited number of trials, the reconstruction quality is superior to RC or STA.

## 2 Mapping synaptic inputs onto one neuron

We start by formalizing the problem of mapping synaptic connections from a population of $N$ potentially pre-synaptic neurons onto a single neuron, as exemplified by granule cells synapsing onto a Purkinje cell (Figure 1a). Our experimental protocol can be illustrated using linear algebra formalism, Figure 1b. We represent synaptic weights as components of a column vector $x$, where zeros represent non-existing connections. Each row in the stimulation matrix $A$ represents a trial, ones indicating neurons driven to spike once and zeros indicating non-spiking neurons. The number of rows in the stimulation matrix $A$ is equal to the number of trials $M$. The column vector $y$ represents $M$ measurements of membrane voltage obtained by an intra-cellular recording from the post-synaptic neuron:

$$y = Ax. \qquad (1)$$

In order to recover individual synaptic weights, Eq. (1) must be solved for $x$. RC (or STA) solution to this problem is $x = (A^T A)^{-1} A^T y$, which minimizes $(y-Ax)^2$ if $M>N$. In the case $M<N$, the corresponding expression $x = A^T(AA^T)^{-1}y$ is a solution to the following problem:

$$\min \|x\|_{l_2} = \sqrt{\sum_{i=1}^{N} x_i^2}, \text{ subject to } y = Ax.$$

Given prior knowledge that the connectivity is sparse, we propose to recover $x$ by solving instead:

$$\min \|x\|_{l_0}, \qquad \text{subject to } y = Ax,$$

where $\|x\|_{l_0}$ is the $l_0$-norm of $x$ or the number of non-zero elements. Under certain conditions

[25-29], this solution can be obtained by minimizing the $l_1$-norm: $\|x\|_{l_1} = \sum_{i=1}^{N} |x_i|$ using linear programming [38] or by iterative greedy algorithms [39, 40]. In this paper, we used a particularly efficient Compressive Sampling Matching Pursuit (CoSaMP) algorithm [41, 42].

We simulate the proposed reconstruction method *in silico* by generating a neural network, simulating experimental measurements, and recovering synaptic weights. We draw unitless synaptic weights from a distribution derived from electrophysiological measurements [4, 5, 43, 44] containing a delta-function at zero and an exponential distribution with a unit mean (Figure 2a). We generate an $M$-by-$N$ stimulation matrix $A$ by setting $F$ randomly chosen entries in each row to one, and the rest to zero. We compute the measurement vector $y$ by multiplying $A$ and $x$. Then, we use the CoSaMP algorithm to recover synaptic weights, $x_r$, from $A$ and $y$. Figure 2b compares a typical result of such reconstruction and a result of RC with originally generated non-zero synaptic weights $x$. Despite using fewer measurements than required in the brute force approach, CS achieves perfect reconstruction while RC yields a worse result [45].

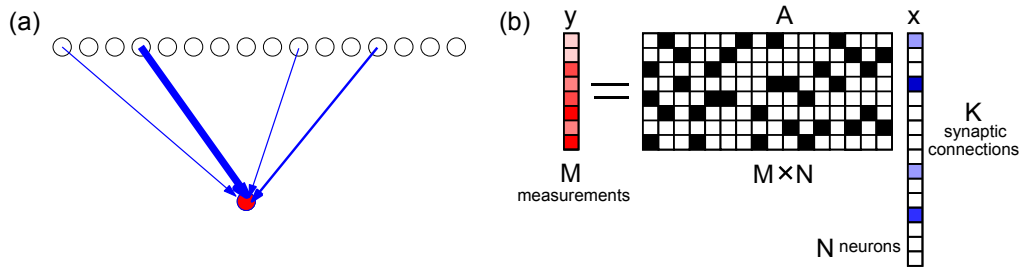

Figure 1: Mapping synapses onto one neuron. a) A potentially post-synaptic neuron (red) receives synaptic connection (blue) from $K$ neurons out of $N$ potentially pre-synaptic neurons. b) Linear algebra representation of the experimental protocol. The column vector $x$ contains synaptic weights from $N$ potentially pre-synaptic neurons: $K$ blue squares represent existing connections, white squares represent absent connections. The matrix $A$ represents the sequence of stimulation: black squares in each row represent stimulated neurons in each trial. The column vector $y$ contains measured membrane voltage in the red neuron.

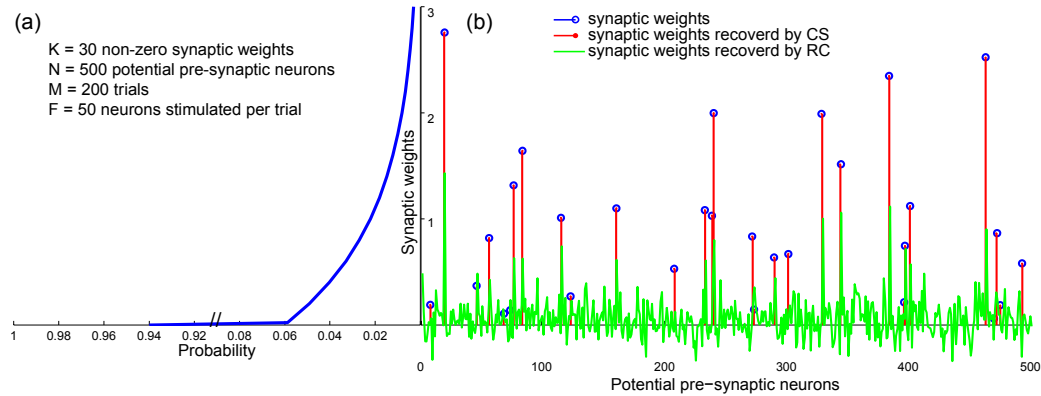

Figure 2: Reconstruction of synaptic weights onto one neuron. a) Synaptic weights are drawn from the empirically motivated probability distribution. b) Reconstruction by CS (red) coincides perfectly with generated synaptic weights (blue), achieving 60% improvement in the number of trials over the brute force approach. RC result (green) is significantly worse.

## 3 Minimum number of measurements as a function of network size and sparseness

In order to understand intuitively why the number of trials can be less than the number of potential synapses, note that the minimum number of trials, i.e. information or entropy, is given by the logarithm of the total number of possible connectivity patterns. If connections are binary, the number of different connectivity patterns onto a post-synaptic neuron from $N$ neurons is $2^N$, and hence the minimum number of trials is $N$. However, prior knowledge that only $K$ connections are present reduces the number of possible connectivity patterns from $2^N$ to the binomial coefficient, $C_K^N \sim (N/K)^K$. Thus, the number of trials dramatically reduces from $N$ to $K \log(N/K) \ll N$ for a sparse circuit.

In this section we search computationally for the minimum number of trials required for exact reconstruction as a function of the number of non-zero synaptic weights $K$ out of $N$ potentially pre-synaptic neurons. First, note that the number of trials depends on the number of stimulated neurons $F$. If $F = 1$ we revert to the brute force approach and the number of measurements is $N$, while for $F = N$, the measurements are redundant and no finite number suffices. As the minimum number of measurements is expected to scale as $K \log N$, there must be an optimal $F$ which makes each measurement most informative about $x$.

To determine the optimal number of stimulated neurons $F$ for given $K$ and $N$, we search for the minimum number of trials $M$, which allows a perfect reconstruction of the synaptic connectivity $x$. For each $F$, we generate 50 synaptic weight vectors and attempt reconstruction from sequentially increasing numbers of trials. The value of $M$, at which all 50 recoveries are successful (up to computer round-off error), estimates the number of trial needed for reconstruction with probability higher than 98%. By repeating this procedure 50 times for each $F$, we estimate the mean and standard deviation of $M$. We find that, for given $N$ and $K$, the minimum number of trials, $M$, as a function of the number of stimulated neurons, $F$, has a shallow minimum. As $K$ decreases, the minimum shifts towards larger $F$ because more neurons should be stimulated simultaneously for sparser $x$. For the explored range of simulation parameters, the minimum is located close to *0.1N*.

Next, we set $F = 0.1N$ and explore how the minimum number of measurements required for exact reconstruction depends on $K$ and $N$. Results of the simulations following the recipe described above are shown in Figure 3a. As expected, when $x$ is sparse, $M$ grows approximately linearly with $K$ (Figure 3b), and logarithmically with $N$ (Figure 3c).

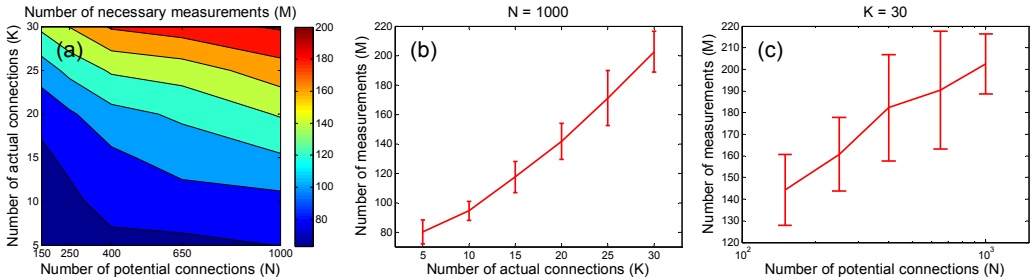

Figure 3: a) Minimum number of measurements $M$ required for reconstruction as a function of the number of actual synapses, $K$, and the number of potential synapses, $N$. b) For given $N$, we find $M \sim K$. c) For given $K$, we find $M \sim \log N$ (note semi-logarithmic scale in c).

## 4    Robustness of reconstructions to noise and violation of simplifying assumptions

To make our simulation more realistic we now take into account three possible sources of noise: 1) In reality, post-synaptic voltage on a given synapse varies from trial to trial [4, 5, 46-52], an effect we call synaptic noise. Such noise detrimentally affects reconstructions because each row of $A$ is multiplied by a different instantiation of vector $x$. 2) Stimulation of neurons may be imprecise exciting a slightly different subset of neurons than intended and/or firing intended neurons multiple times. We call this effect stimulation noise. Such noise detrimentally affects reconstructions because, in its presence, the actual measurement matrix $A$ is different from the one used for recovery. 3) A synapse may fail to release neuro-transmitter with some probability.

Naturally, in the presence of noise, reconstructions cannot be exact. We quantify the reconstruction error by the normalized $l_2$-error $\|x-x_r\|_{l_2}/\|x\|_{l_2}$, where $\|x-x_r\|_{l_2}=\sqrt{\sum_{i=1}^{N}(x_i-x_{ri})^2}$. We plot normalized reconstruction error in brute force approach ($M = N = 500$ trials) as a function of noise, as well as CS and RC reconstruction errors ($M = 200, 600$ trials), Figure 4.

For each noise source, the reconstruction error of the brute force approach can be achieved with 60% fewer trials by CS method for the above parameters (Figure 4). For the same number of trials, RC method performs worse. Naturally, the reconstruction error decreases with the number of trials. The reconstruction error is most sensitive to stimulation noise and least sensitive to synaptic noise.

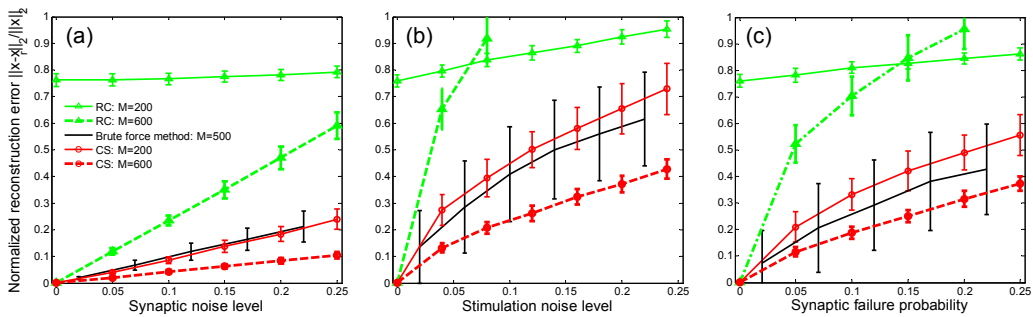

Figure 4: Impact of noise on the reconstruction quality for $N = 500$, $K = 30$, $F = 50$. a) Recovery error due to trial-to-trial variation in synaptic weight. The response $y$ is calculated using the synaptic connectivity $x$ perturbed by an additive Gaussian noise. The noise level is given by the coefficient of variation of synaptic weight. b) Recovery error due to stimulation noise. The matrix $A$ used for recovery is obtained from the binary matrix used to calculate the measurement vector $y$ by shifting, in each row, a fraction of ones specified by the noise level to random positions. c) Recovery error due to synaptic failures.

The detrimental effect of the stimulation noise on the reconstruction can be eliminated by monitoring spiking activity of potentially pre-synaptic neurons. By using calcium imaging [18-22], voltage-sensitive dyes [23] or multi-electrode arrays [17, 26] one could record the actual stimulation matrix. Because most random matrices satisfy the reconstruction requirements [31, 34, 35], the actual stimulation matrix can be used for a successful recovery instead of the intended one.

If neuronal activity can be monitored reliably, experiments can be done in a different mode altogether. Instead of stimulating designated neurons with high fidelity by using highly localized and intense light, one could stimulate all neurons with low probability. Random firing events can be detected and used in the recovery process. The light intensity can be tuned to stimulate the optimal number of neurons per trial.

Next, we explore the sensitivity of the proposed reconstruction method to the violation of simplifying assumptions. First, whereas our simulation assumes that the actual number of connections, $K$, is known, in reality, connectivity sparseness is known *a priori* only approximately. Will this affect reconstruction results? In principle, CS does not require prior knowledge of $K$ for reconstruction [31, 34, 35]. For the CoSaMP algorithm, however, it is important to provide value $K$ larger than the actual value (Figure 5a). Then, the algorithm will find all the actual synaptic weights plus some extra non-zero weights, negligibly small when compared to actual ones. Thus, one can provide the algorithm with the value of $K$ safely larger than the actual one and then threshold the reconstruction result according to the synaptic noise level.

Second, whereas we assumed a linear summation of inputs [53], synaptic integration may be non-linear [54]. We model non-linearity by setting $y = y_l + \alpha y_l^2$, where $y_l$ represents linearly summed synaptic inputs. Results of simulations (Figure 5b) show that although non-linearity can significantly degrade CS reconstruction quality, it still performs better than RC.

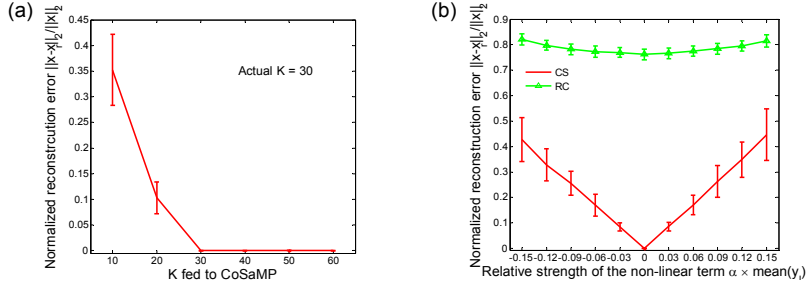

Figure 5: Sensitivity of reconstruction error to the violation of simplifying assumptions for $N = 500$, $K = 30$, $M = 200$, $F = 50$. a) The quality of the reconstruction is not affected if the CoSaMP algorithm is fed with the value of $K$ larger than actual. b) Reconstruction error computed in 100 realizations for each value of the quadratic term relative to the linear term.

## 5  Mapping synaptic inputs onto a neuronal population

Until now, we considered reconstruction of synaptic inputs onto one neuron using sub-threshold measurements of its membrane potential. In this section, we apply CS to reconstructing synaptic connections onto a population of potentially post-synaptic neurons. Because in CS the choice of stimulated neurons is non-adaptive, by recording from all potentially post-synaptic neurons in response to one sequence of trials one can reconstruct a complete feedforward network (Figure 6).

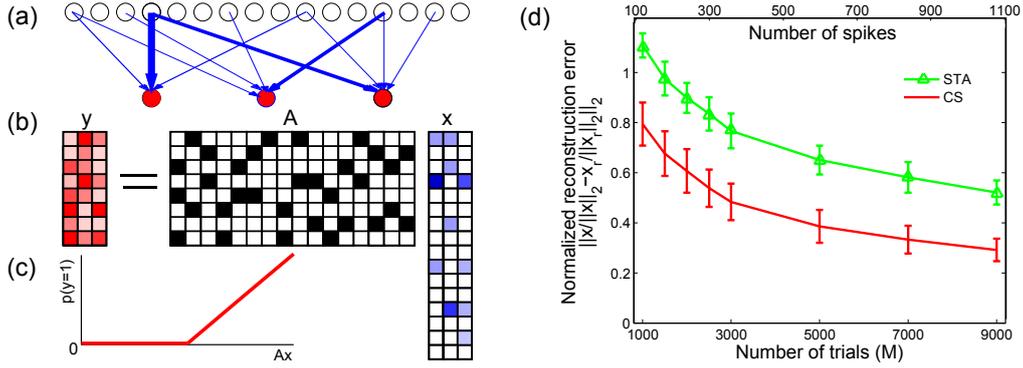

Figure 6: Mapping of a complete feedforward network. a) Each post-synaptic neuron (red) receives synapses from a sparse subset of potentially pre-synaptic neurons (blue). b) Linear algebra representation of the experimental protocol. c) Probability of firing as a function of synaptic current. d) Comparison of CS and STA reconstruction error using spike trains for $N = 500$, $K = 30$ and $F = 50$.

Although attractive, such parallelization raises several issues. First, patching a large number of neurons is unrealistic and, therefore, monitoring membrane potential requires using different methods, such as calcium imaging [18-22], voltage sensitive dyes [23-25] or multi-electrode arrays [17, 26]. As these methods can report reliably only spiking activity, the measurement is not analog but discrete. Depending on the strength of summed synaptic inputs compared to the firing threshold, the postsynaptic neuron may be silent, fire once or multiple times. As a result, the measured response $y$ is quantized by the integer number of spikes. Such quantized measurements are less informative than analog measurements of the sub-threshold membrane potential. In the extreme case of only two quantization levels, spike or no spike, each measurement contains only 1 bit of information. Therefore, to achieve reasonable reconstruction quality using quantized measurements, a larger number of trials $M >> N$ is required.

We simulate circuit reconstruction from spike recordings *in silico* as follows. First, we draw synaptic weights from an experimentally motivated distribution. Second, we generate a random stimulation matrix and calculate the product $Ax$. Third, we linear half-wave rectify this product and use the result as the instantaneous firing rate for the Poisson spike generator (Figure 6c). We used a rectifying threshold that results in *10%* of spiking trials as typically observed in experiments. Fourth, we reconstruct synaptic weights using STA and CS and compare the results with the generated weights. We calculated mean error over *100* realizations of the simulation protocol (Figure 6d).

Due to the non-linear spike generating procedure, $x$ can be recovered only up to a scaling factor. We propose to calibrate $x$ with a few brute-force measurements of synaptic weights. Thus, in calculating the reconstruction error using $l_2$ norm, we normalize both the generated and recovered synaptic weights. Such definition is equivalent to the angular error, which is often used to evaluate the performance of STA in mapping receptive field [37, 55].

Why is CS superior to STA for a given number of trials (Figure 6d)? Note that spikeless trials, which typically constitute a majority, also carry information about connectivity. While STA discards these trials, CS takes them into account. In particular, CoSaMP starts with the STA solution as zeroth iteration and improves on it by using the results of all trials and the sparseness prior.

# 6 Discussion

We have demonstrated that sparse feedforward networks can be reconstructed by stimulating multiple potentially pre-synaptic neurons simultaneously and monitoring either sub-threshold or spiking response of potentially post-synaptic neurons. When sub-threshold voltage is recorded, significantly fewer measurements are required than in the brute force approach. Although our method is sensitive to noise (with stimulation noise worse than synapse noise), it is no less robust than the brute force approach or RC.

The proposed reconstruction method can also recover inputs onto a neuron from spike counts, albeit with more trials than from sub-threshold potential measurements. This is particularly useful when intra-cellular recordings are not feasible and only spiking can be detected reliably, for example, when mapping synaptic inputs onto multiple neurons in parallel. For a given number of trials, our method yields smaller error than STA.

The proposed reconstruction method assumes linear summation of synaptic inputs (both excitatory and inhibitory) and is sensitive to non-linearity of synaptic integration. Therefore, it is most useful for studying connections onto neurons, in which synaptic integration is close to linear. On the other hand, multi-neuron stimulation is closer than single-neuron stimulation to the intrinsic activity in the live brain and can be used to study synaptic integration under realistic conditions.

In contrast to circuit reconstruction using intrinsic neuronal activity [56, 57], our method relies on extrinsic stimulation of neurons. Can our method use intrinsic neuronal activity instead? We see two major drawbacks of such approach. First, activity of non-monitored pre-synaptic neurons may significantly distort reconstruction results. Thus, successful reconstruction would require monitoring all active pre-synaptic neurons, which is rather challenging. Second, reliable reconstruction is possible only when the activity of pre-synaptic neurons is uncorrelated. Yet, their activity may be correlated, for example, due to common input.

We thank Ashok Veeraraghavan for introducing us to CS, Anthony Leonardo for making a retina dataset available for the analysis, Lou Scheffer and Hong Young Noh for commenting on the manuscript and anonymous reviewers for helpful suggestions.